\documentclass[aps,pra,superscriptaddress,floatfix,showpacs,10pt,twocolumn]{revtex4}
\usepackage[dvips]{graphicx}
\usepackage{amsmath}
\begin{document}
\title{Entanglement swapping without joint Measurement}
\author{Ming Yang}
\email{mingyang@ahu.edu.cn}
\author{Wei Song}
\author{Zhuo-Liang Cao}
\email{zlcao@ahu.edu.cn} \affiliation{Anhui Key Laboratory of
Information Material {\&} Devices,School of Physics {\&} Material Science, \\
Anhui University, Hefei, 230039, PRChina}

\begin{abstract}
We propose an entanglement swapping scheme in cavity QED. In the
scheme, the previously used joint measurement is not needed. The
entanglement swapping in our proposal is a non-post-selection one,
i.e., after the swapping is done, the swapped entanglement is
still there.
\end{abstract}

\maketitle

Entanglement plays an important role in quantum information processing
(QIP). Several schemes have been proposed to generate entangled states, such
as the nonlinear interaction between optical pulse and nonlinear crystal~%
\cite{photon}, the interaction between different particles and so on~\cite%
{atom1, atom2}. After the entanglement generation, the entangled
particles must be distributed among distant users for quantum
communication purpose. During the distribution process, the
entanglement of the particles will inevitably decrease. The longer
the distance, the bigger the decrease. To avoid this problem, an
alternative method has been proposed to generate entanglement
between two distant particles that have never interacted before.
That is the so-called entanglement swapping~\cite{swapp}. In this
method, there are usually three spatially separate users, and two
of them have shared one pair of entangled particles with the
third. Then the third user will operate a joint measurement (such
as Bell state measurement) on the two particles he possesses.
Corresponding to the measurement result, the two particles
possessed by the two spatially separate users will collapse into
an entangled state without any entanglement before the joint
measurement. Recently, entanglement swapping schemes have been
proposed using linear optical elements with post-selection
measurement~\cite{pan} or without post-selection
measurement~\cite{wang}. Besides the entanglement generation,
entanglement purification is another application of entanglement swapping~%
\cite{bose}. So the realization of entanglement swapping is very
important for the quantum communication. Because the Bell states
measurement or other types of joint measurement is the core of the
previous entanglement swapping schemes, the realization of Bell
state measurement is of the key importance for the entanglement
swapping. Hitherto, we still can not discriminate the four Bell
states totally and conclusively. At the same time, the realization
of the Bell state measurement is still difficult in experiment,
especially for the atomic Bell states.

To overcome this difficulty, we propose a new entanglement
swapping scheme, where we only need single measurement rather than
joint measurement. Although, our scheme follows Zheng's
teleportation scheme~\cite{zheng}, where he realized the
teleportation of unknown atomic state without Bell states
measurement, we have improved the scheme. In this paper, we
discuss a scheme for entanglement swapping using concepts of
cavity QED. Initially, one party A shares an entangled pair of
atoms with another party C. C also shares an entangled pair of
cavities with a third party B. Next, the atom and the cavity of C
are made to interact for a fixed time after which a
state-measurement is performed on atom or on cavity. For both
cases it is shown that, with a certain finite probability, the
final state of A's atom and B's cavity is a maximally entangled
state.

Although the creation of entangled photons is relatively easy when
compared with entangling cavities and atoms, the entangled states
with photons are difficult to be stored for future use. We may
choose to swap the entanglement to two atoms which can be easily
stored for future use. This has potential value for real
application in practice in the future.

Next, let's go into the detailed entanglement swapping scheme.
Suppose there are three spatially separate users Alice, Bob and
Clare. Alice and Clare have shared a pair of atoms ($1$, $2$) with
atom $1$ belonging to Alice, atom $2$ belonging to Clare. These
two atoms have been previously prepared in the following entangled
state:
\begin{equation}
\label{eq1}
|\Phi\rangle_{12}=a|e\rangle_{1}|e\rangle_{2}+b|g\rangle_{1}|g\rangle_{2},
\end{equation}
where $a$ and $b$ are the normalization coefficients.

Besides the atom $2$, Clare also possesses one single mode cavity $3$, which
is entangled with another single mode cavity $4$, and the cavity $4$ belongs
to Bob. Similarly, we also suppose the two cavities have been prepared in
the following entangled state:
\begin{equation}
\label{eq2}
|\Phi\rangle_{34}=a|1\rangle_{3}|1\rangle_{4}+b|0\rangle_{3}|0\rangle_{4},
\end{equation}
where $|0\rangle$ and $|1\rangle$ denote the vacuum state and one
photon state of the cavity mode respectively, and the
normalization coefficients are all the same to the atomic
entangled state in Eq(\ref{eq1}). From the above two entangled
states, we conclude that there is no correlation between the atom
$1$ and cavity $4$ at this moment. After entanglement swapping,
the atom $1$ and cavity $4$ will be left in a maximally entangled
state.

To realize the entanglement swapping, Clare will let the atom $2$ through
the cavity $3$. Suppose the atomic transition frequency is resonant with the
cavity mode, then, in the interaction picture the interaction can be
described as:
\begin{equation}
\label{eq3} H_{I}=g(aS^{+}+a^{+}S^{-}),
\end{equation}
where $g$ is the coupling constant between the atom and the cavity
mode, $a$ and $a^{+}$ are annihilation and creation operators of
the cavity mode respectively, and $S^{+}=|e\rangle\langle g|,$
$S^{-}=|g\rangle\langle  e|$ are raising
and lowering operators for atom $2$ with $|e\rangle ,$ $%
|g\rangle $ being the excited state and ground state of the atoms
respectively.

Before swapping, the state of the total system is:
\begin{align}
\label{eq4} |\Psi\rangle&_{1234}=\nonumber\\
&(a|e\rangle_{1}|e\rangle_{2}+b|g\rangle_{1}|g\rangle_{2})\otimes(a|1\rangle_{3}|1\rangle_{4}+b|0\rangle_{3}|0\rangle_{4}).
\end{align}

After interaction time $t$, the state of the total system will
evolve into the following state:
\begin{align}\label{eq5}
 |\Psi\rangle^{'}&_{1234}=\nonumber\\
 & a^{2}|e\rangle_{1}|1\rangle_{4}[cos(\sqrt{2}gt)|e\rangle_{2}|1\rangle_{3}-isin(\sqrt{2}gt)|g\rangle_{2}|2\rangle_{3}]\nonumber\\
 & +ab|e\rangle_{1}|0\rangle_{4}[cos(gt)|e\rangle_{2}|0\rangle_{3}-isin(gt)|g\rangle_{2}|1\rangle_{3}]\nonumber\\
 & +ab|g\rangle_{1}|1\rangle_{4}[cos(gt)|g\rangle_{2}|1\rangle_{3}-isin(gt)|e\rangle_{2}|0\rangle_{3}]\nonumber\\
 & +b^{2}|g\rangle_{1}|0\rangle_{4}|g\rangle_{2}|0\rangle_{3}.
\end{align}

After the atom flying out of the cavity, Clare will detect the atom $2$. If
the atom $2$ is detected in excited state, the atom $1$ and cavities $3$, $4$
will collapse into:
\begin{align}
\label{eq6}
|\Psi\rangle^{'}_{134} =& N\{ab[cos(gt)|e\rangle_{1}|0\rangle_{4}-isin(gt)|g\rangle_{1}|1\rangle_{4}]|0\rangle_{3}\nonumber\\
  &+a^{2}cos(\sqrt{2}gt)|e\rangle_{1}|1\rangle_{3}|1\rangle_{4}\},
\end{align}
where $N$ is the normalization factor. If Clare chooses the
interaction time to satisfy $gt=7\pi /4$, we get that $\cos
\sqrt{2}gt=0.079\approx 0$. So the
third term in Eq(\ref{eq6}) can be eliminated. Then the atom $1$ and cavity $%
4$ collapsed approximately into a maximally entangled state without
detection on the cavity $3$:
\begin{equation}
\label{eq7}
|\Psi\rangle_{14}=\frac{1}{\sqrt{2}}(|e\rangle_{1}|0\rangle_{4}+i|g\rangle_{1}|1\rangle_{4}),
\end{equation}
with probability $P=|b|^{2}\times(1-|b|^{2})$. After a rotation
operation, the entangled state can be transformed into the
standard form with a zero relative phase factor. The fidelity of
the output state relative to a perfect maximally entangled state
is $F={|b|^{2}}/{[|b|^{2}+(1-|b|^{2})\cos^{2}{(
{\sqrt{2}gt})}]}$. From the Fig.1, we conclude that the fidelity is bigger than $0.9$ unless $%
b<0.25$, and the fidelity reaches $0.99$ when $b=0.6$. The biggest
successful probability can reach $0.25$. After swapping, Bob can
send a resonant atom through the cavity $4$. If Bob sets the
interaction time appropriately, the interaction can swap the atom
and cavity excitations. Then the two atoms, which belong to Alice
and Bob, and never interact before, are in a maximally entangled
state.

\begin{figure}[htbp]
\includegraphics[scale=0.47]{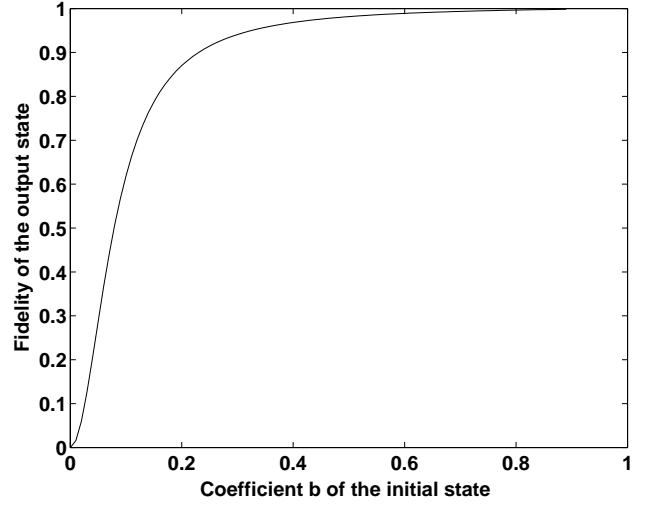}
\caption{The fidelity of the output state relative to a perfect
maximally entangled state varies with the coefficient b of the
initial state. Here $gt=7\pi/4$.} \label{fig1}
\end{figure}

From the above process, we found that the measurement on atomic
state is still needed in the above scheme. Next, we will prove
that this scheme also can realize the entanglement swapping
without the measurement on atoms. From the evolution result in
Eq(\ref{eq5}), if the coefficient b of the initial state is
relative small, the last term of Eq(\ref{eq5}) can be eliminated
from the total state approximately. Then Clare can detect the
cavity $3$ rather than the atom $2$ to complete the entanglement
swapping process. Further more, Clare only need an ordinary photon
detector, which only distinguishes the vacuum and non-vacuum Fock
states, rather than a sophisticated photon detector that can
distinguish one and two photons. If the cavity $3$ is detected in
vacuum state, the atoms $1$, $2$ and cavity $4$ will collapse
into:
\begin{align}\label{eq8}
 |\Psi\rangle^{'}_{124} =& N^{'}\{ab[cos(gt)|e\rangle_{1}|0\rangle_{4}-isin(gt)|g\rangle_{1}|1\rangle_{4}]|e\rangle_{2}\nonumber\\
  &+b^{2}|g\rangle_{1}|0\rangle_{4}|g\rangle_{2}\},
\end{align}
where ${N}^{\prime }$ is the normalization factor. If Clare
chooses the interaction time to satisfy $gt=7\pi /4$, the atom $1$
and cavity $4$ will collapse approximately into a maximally entangled state expressed in Eq (%
\ref{eq7}) without detection on the atom $2$. The probability of
obtaining this state is still
${P}^{\prime}=|b|^{2}\times(1-|b|^{2})$, and the fidelity of the
output state relative to the perfect maximally entangled state is
${F}^{\prime }=1-|b|^{2}$. To achieve ${F}^{\prime}=0.96$, the
coefficient $b$ must be $0.2$, which still guarantee that the
fidelity is bigger than $0.95$. Here, the successful probability
is $0.04$.

The coefficients in Eqs. (\ref{eq1}) and (\ref{eq2}) are chosen to be
identical in this paper (in amplitude and phase). If there exists an error
between the coefficients in Eqs. (\ref{eq1}) and (\ref{eq2}), what will the
result become? Through analysis, we find that the current scheme allows the
existence of a small error between the coefficients in Eqs. (\ref{eq1}) and (%
\ref{eq2}). Suppose the states in Eqs (\ref{eq1}) and (\ref{eq2})
can be reexpressed as:

\begin{equation}\label{eq9}
|\Phi\rangle_{12}=\sqrt{1-b^{2}}|e\rangle_{1}|e\rangle_{2}+b|g\rangle_{1}|g\rangle_{2},
\end{equation}
\begin{equation}\label{eq10}
|\Phi\rangle_{34}=\sqrt{1-b^{2}(1+k)^{2}}|1\rangle_{3}|1\rangle_{4}+b(1+k)|0\rangle_{3}|0\rangle_{4},
\end{equation}
where $k$ is the small error rate constant of the coefficient $b$, and $k$, $%
b$ must satisfy $|b|<1, |b(1+k)|<1$. To deduce the success
probability and the fidelity of the output state, we will consider
the first case as example. If the above mentioned error exists,
the success probability and the fidelity of the output state will
become:

\begin{equation}
P_{new}=\frac{1}{2}\left\{{\left( {1-b^{2}}\right)b^{2}(1+k)^{2}+b^{2}}%
\left[{1-b^{2}(1+k)^{2}}\right] \right\} \label{eq11}
\end{equation}
\begin{widetext}
\begin{equation}
F_{new}=\frac{{\frac{1}{2}}\left[ {\sqrt{1-b^{2}}\,b(1+k)+b\sqrt{
1-b^{2}(1+k)^{2}}}\right] ^{2}}{\left( {1-b^{2}}\right)
\,b^{2}(1+k)^{2}+b^{2}\left[ {1-b^{2}(1+k)^{2}}\right] +2\left(
{1-b^{2}} \right) \,\left[ {1-b^{2}(1+k)^{2}}\right] \cos
^{2}\sqrt{2}gt}  \label{eq12}
\end{equation}
\end{widetext}

If we let the error rate constant $k=0.1$, $b=0.6$, $P_{new}=0.24098$, $%
F_{new}=0.98463$. That is to say, this kind of error will affect the
probability of success and the fidelity very slightly. The same analysis
applies to the second case.

In addition, the current scheme requires the entangled pairs of
atoms (cavities) be distributed between A and C (B and C), which
should be separated by large distances. This point can not easily
be realized within the current cavity QED technology. Unlike the
photon-based schemes, where optical fibers provide a low-loss
communication channel, sending single atoms (or entire cavities)
over large distances while maintaining their quantum state is
still beyond present technology. This may narrow the application
of the current scheme. But swapping and teleportation are not only
supposed to make quantum communication between two remote parties,
it has broad applications in quantum information processing, e.g.,
in quantum computation~\cite{chuang}. We mean, even though the two
cavities are not separated far away, the swapping is still useful
in QIP. For another example, the entangled states with photons are
difficult to be stored for future use. We may choose to swap the
entanglement to two atoms which can be easily stored for future
use.

Next, we will discuss the feasibility of the scheme. From reference~\cite%
{zheng, experiment1}, we get that, we should make use of Rydberg
atoms with
principal quantum numbers 50 and 51, because their radiative time are $%
T_{r}=3\times 10^{-2}s$. For a normal cavity, the decay time can reach $%
T_{c}=1.0\times 10^{-3}s$. The coupling constant is $g=2\pi \times 25kHz$.
Then we get that the interaction time of atom and cavity is ${7\pi }%
/4g=3.5\times 10^{-5}s$, so we can evaluate that the total time for the
whole scheme is about $T=3.5\times 10^{-4}s$, which is much shorter than $%
T_{r},\;T_{c}$. Hence, the current scheme might be realizable in
the near future.

In the paper, we considered the initial states of the form
$a|e\rangle_{1}|e\rangle_{2}+b|g\rangle_{1}|g\rangle_{2}$ for
atoms and
$a|1\rangle_{3}|1\rangle_{4}+b|0\rangle_{3}|0\rangle_{4}$ for
cavities just for clarifying the principle of the current scheme.
Considering the feasibility of the scheme, initial states of the
form $a|g\rangle_{1}|e\rangle_{2}+b|e\rangle_{1}|g\rangle_{2}$ for
atoms and
$a|1\rangle_{3}|0\rangle_{4}+b|0\rangle_{3}|1\rangle_{4}$ for
cavities might be more realistic, and this kind of entangled
states for Rydberg atoms and microwave cavities have been prepared
by Haroche's group in experiment~\cite{experiment1, experiment2}.
Further calculation suggests that initial states of the form
$a|g\rangle_{1}|e\rangle_{2}+b|e\rangle_{1}|g\rangle_{2}$ for
atoms and
$a|1\rangle_{3}|0\rangle_{4}+b|0\rangle_{3}|1\rangle_{4}$ for
cavities also can lead to the same results as derived in the
current scheme.

Seemingly only one measurement is needed in the schemes, but a
coincidence measurement (of one atom and one cavity) is
essentially required to obtain a maximally entangled state as
in~\cite{pan}. Generally, a single measurement never yields ideal
output. Here because the state to be measured can be approximately
factorized from the total state of the system, we replace the
coincidence measurement with a single measurement.

Based on cavity QED, we have presented an entanglement swapping
scheme. The most distinct advantage of it is that it does not need
a joint measurement needed by the previous entanglement swapping
schemes. It only needs a resonant interaction between an atom and
a cavity mode and a measurement on cavity (or atom). Our proposal
is for non-post-selection, i.e., after the swapping is done, the
swapped entanglement is still there. This has potential value for
real application in practice in the future.

\begin{acknowledgments}
We gratefully acknowledge Dr Xiang-Bin Wang's helpful suggestions,
and we appreciate the challenging questions raised by the
anonymous referee. This work is supported by the Key Program of
the Education Department of Anhui Province under Grant
No:2004kj005zd, Anhui Provincial Natural Science Foundation under
Grant No: 03042401 and the Talent Foundation of Anhui University.
\end{acknowledgments}

\end{document}